\documentclass{ecai} 

\usepackage{url}
\usepackage{mdframed}

\usepackage{amsmath,amsthm}
\usepackage{amsfonts}
\usepackage{mathtools}
\usepackage{xspace}
\usepackage{relsize}
\usepackage{stmaryrd}
\usepackage{multirow}
\usepackage{makecell}
\usepackage{hyperref}
\usepackage{tikz}
\usetikzlibrary{shapes,arrows,backgrounds,positioning,patterns,fit}
\usetikzlibrary{calc,automata,graphs,chains,trees}
\tikzset{
->, 
>=stealth', 
node distance=2cm, 
initial text=$ $, 
}

\usepackage{balance}

\newtheorem{example}{Example}
\newtheorem{theorem}{Theorem}
\newtheorem{lemma}{Lemma}
\newtheorem{definition}{Definition}

\newcommand{\mst}{\mathit{st}}

\newcommand{\RV}{\mathit{RV}}

\newcommand{\PS}{\textsc{PS}}

\newcommand{\CS}{\textsc{CS}}

\newcommand{\CV}{\textsc{CV}}

\newcommand{\PV}{\textsc{PV}}

\newcommand{\field}{\mathit{id}}

\newcommand{\LTL}{\ensuremath{\text{LTL}}}
\newcommand{\LTLf}{\ensuremath{\text{LTL}_f}}
\newcommand{\LTLMT}{\ensuremath{\text{LTL}_f^\textrm{MT}}}
\newcommand{\toltl}{\mathit{abs}}

\newcommand{\Next}{\bigcirc}
\DeclareRobustCommand{\WNext}{{\mathrlap{\bigcirc}\!\sim}}

\newcommand{\sem}[1]{\llbracket{#1}\rrbracket}
\newcommand{\err}{\nabla}
\newcommand{\Werr}{\widetilde{\nabla}}

\newcommand{\true}{\mathit{true}}
\newcommand{\false}{\mathit{false}}

\newcommand{\Until}{U}
\newcommand{\Tom}{X}
\newcommand{\WTom}{\widetilde{X}}

\newcommand{\SDWA}{{\sc Sdwa}\xspace}

\newcommand{\dsig}{\mathcal{S}}
\newcommand{\dtheory}{\mathcal{D}}
\newcommand{\aut}{\mathcal{A}}

\newcommand{\bool}{\mathbb{B}}
\newcommand{\run}{\pi}
\newcommand{\CHC}{\mbox{CHC}}
\newcommand{\AP}{\text{AP}}

\newcommand{\MSOD}{\mbox{\sc Mso-D}\xspace}

\newcommand*{\defeq}{\stackrel{\mathsmaller{\mathsf{def}}}{=}}

\newcommand{\prop}{b}

\newcommand{\rel}{p}
\newcommand{\langW}{\mathcal{L}}
\newcommand{\langE}{\mathcal{E}}

\newcommand{\theoryconsts}
{\mathcal{D}^\mathrm{con}}

\newcommand{\theorypreds}
{\mathcal{D}^\mathrm{rel}}

\newcommand{\theoryfuns}
{\mathcal{D}^\mathrm{fun}}

\newcommand{\temp}{\mathit{temp}}

\renewcommand{\phi}{\varphi}
\newcommand{\tempctrl}{\textsc{TempCtrl}}
\newcommand{\LIA}{\textsc{Lia}}
\newcommand{\LRA}{\textsc{Lra}}
\newcommand{\imposs}{\textsc{GandF}}

\begin{document}

\begin{frontmatter}
\paperid{1256}

\title{A Unified Automata-Theoretic Approach to {\LTLf} Modulo Theories (Extended Version)}


\author[A]{\fnms{Marco}~\snm{Faella}\orcid{0000-0001-7617-5489}\thanks{Corresponding Author. Email: m.faella@unina.it.}\footnote{Equal contribution.}}
\author[B]{\fnms{Gennaro}~\snm{Parlato}\orcid{0000-0002-8697-2980}\thanks{Corresponding Author. Email: gennaro.parlato@unimol.it.}\footnotemark}
 
\address[A]{Univeristy of Naples, Federico II, Italy}
\address[B]{University of Molise, Italy}

\begin{abstract}
We present a novel automata-based approach to address linear temporal logic modulo theory (\LTLMT) as a specification language for data words.
{\LTLMT} extends {\LTLf} by replacing atomic propositions with quantifier-free multi-sorted first-order formulas interpreted over arbitrary theories. 
While standard  {\LTLf} is reduced to finite automata,
we reduce {\LTLMT} to symbolic data-word automata ({\SDWA}s),
whose transitions are guarded by constraints from underlying theories.
Both the satisfiability of {\LTLMT} and the emptiness of {\SDWA}s are undecidable, but the latter can be reduced to a system of constrained Horn clauses, which are supported by efficient 
solvers and ongoing research efforts.
We discuss multiple applications of our approach beyond satisfiability, 
including model checking and runtime monitoring.
Finally, a set of empirical experiments shows that our approach to satisfiability works at least as well as a previous custom 
solution.
\end{abstract}

\end{frontmatter}

\section{Introduction}


Linear Temporal Logic (LTL), pioneered by Pnueli in 1977~\cite{DBLP:conf/focs/Pnueli77},
provides a powerful framework for analyzing temporal properties of dynamic systems, including computer programs, hardware, and protocols. Its
applications include model checking and synthesis of digital circuitry, program verification, cybersecurity, and robotics.
While standard LTL assumes an infinite sequence of observations (which is not always appropriate), \LTLf, introduced by De Giacomo and Vardi in 2013~\cite{DBLP:conf/ijcai/GiacomoV13}, focuses on finite traces instead. Since then, \LTLf\ has gained prominence in AI and business process modeling~\cite{DBLP:conf/bpm/GiacomoMGMM14}.

Both LTL and \LTLf\ are tailored for reasoning about system behaviors defined using propositional logic formulas with Boolean variables—representing states or events. They cannot natively handle systems with data, whose states may involve variables from complex domains like integers or real numbers. To address this limitation, more expressive logics, such as first-order theories, may be required to specify properties of real-world systems operating on data. 


We consider {\LTLMT}, an extension of {\LTLf} where atomic propositions are replaced by quantifier-free multi-sorted first-order formulas interpreted over arbitrary theories. This extension is a fragment of the logic introduced by Geatti et al.~\cite{DBLP:conf/ijcai/GeattiGG22} and aligns with a recent effort on extending Monadic Second Order (MSO) logic to data trees~\cite{DBLP:conf/cav/FaellaP22}. {\LTLMT} can express temporal properties of infinite-state systems described by numerical variables. Notably, it allows the specification of relationships between the current and next values of a state variable, a powerful feature that renders the logic undecidable. For instance, the formula $G(\WNext\,\, x = x+1)$ signifies the constant increase of the variable $x$ at each time step until the end of the finite trace.

We present a \emph{unifying and general framework} to tackle various problems related to {\LTLMT} using symbolic data-word automata ({\SDWA}s) and on constrained Horn clauses ({\CHC}s). 
{\SDWA}s extend finite automata (FAs) by equipping states and alphabet with values taken from possibly infinite domains, and their transitions are guarded by formulas taken from quantifier-free first-order theories.
Notably, the emptiness of {\SDWA}s can be reduced
to the satisfiability of {\CHC}s, a type of logic 
that has proven successful in program verification~\cite{Bjorner2015}.
Efficient algorithmic solutions and tools for {\CHC}s continue to improv, as witnessed by the competition {\sc CHC-COMP}~\cite{DBLP:journals/corr/abs-2109-04635}. Our reduction to {\SDWA}s, akin to the classical automata-theoretic approaches to linear temporal logic, enables us to address diverse decision problems for \LTLMT.


First, we handle the \emph{satisfiability} problem for {\LTLMT}, an undecidable problem that can be reduced to the emptiness problem for {\SDWA}s, and subsequently to the satisfiability problem of {\CHC}s through the previously mentioned constructions. This enables the utilization of efficient procedures offered by various solvers, including the SMT solver Z3~\cite{DBLP:conf/tacas/MouraB08}. We implemented our approach in a prototype tool and used it to assess the satisfiability of a benchmark set, which includes cases used in~\cite{DBLP:conf/ijcai/GeattiGG22} and our running example.
Geatti et al.~\cite{DBLP:conf/ijcai/GeattiGG22} exclusively concentrate on satisfiability, presenting a sound and incomplete procedure for {\LTLMT}. They employ a tableau originally introduced for {\LTL} by Reynolds \cite{Reynolds_2016}, iteratively unfolding for an increasing number of steps. Its status is checked at each iteration using an SMT solver. While their approach adeptly handles the logic's time dimension, it lacks the capability to exploit the fixpoint features of modern {\CHC} solvers. Consequently, contradictory temporal requests cannot be detected without a fixed horizon, as illustrated by the formula $\imposs$ in Section~\ref{sec:experiments}.
Overall, our experiments demonstrate that our approach handles all benchmarks belonging to our fragment, showcasing similar or superior performance compared to the tool implementing the approach of \cite{DBLP:conf/ijcai/GeattiGG22}.

As a second application of our framework, we delve into the \emph{model-checking} problem for infinite state-transition systems against {\LTLMT} specifications. Remarkably, to the best of our knowledge, this problem has not been explored in the existing literature. Leveraging the closure of {\SDWA}s under intersection, we can model the state-transition system itself as an {\SDWA}, allowing us to intersect it with the \SDWA derived from the negation of the specification, and subsequently check its emptiness. Given that the emptiness problem is addressed through {\CHC} solving, we can again employ this efficient technology to resolve the model-checking problem, marking a significant advancement in this unexplored domain.

In the final application of our automata-based framework, we focus on the implementation of {\em runtime monitors} for {\LTLMT} specifications. These monitors serve to evaluate the correctness of a specific run based on log files or on-the-fly assessments. Our solution provides both preliminary and final verdicts, where a final verdict indicates that no change in the evaluation of the specification can occur.
Such early termination (a.k.a.\ \emph{anticipatory} monitoring~\cite{Kallwies23}) is clearly desirable for performance reasons.
Our approach involves tracking the state of the {\SDWA} corresponding to the given specification and solving two {\CHC} instances derived from 
the current state of the monitor. 
We also suggest more scalable solutions by employing over-approximations, albeit with slightly reduced precision.
%

Our technical contributions can be summarized as follows:
\begin{itemize}
\item We enhance a previously established automata construction designed for {\LTLMT} monitoring and linear arithmetic (refer to~\cite{DBLP:conf/aaai/FelliMPW23}) by incorporating an off-the-shelf {\LTLf} procedure and tool, specifically SPOT~\cite{DBLP:conf/cav/Duret-LutzRCRAS22}.
\item We introduce {\SDWA}s as a state-based representation of the syntactically convenient {\LTLMT}, and demonstrate its versatility across multiple applications.
\item  We leverage the connection between {\SDWA}s and {\CHC}s, illustrating that modern {\CHC} solvers can effectively address fundamental problems related to {\LTLMT}.
\item We present a series of satisfiability experiments showing that the {\CHC}-based approach is simpler and more general than a previous custom approach (see~\cite{DBLP:conf/ijcai/GeattiGG22}) but also exhibits comparable efficiency.
\end{itemize}

\section{Data Words}\label{sec: Data Words}
In this section, we introduce data words that we use to represent executions involving variables with unbounded values.

Given two integers $i,j$, with $i\leq j$, we denote by $[i,j]$ the set of integers $k$ satisfying $i\leq k \leq j$, and by $[j]$ the set $[1,j]$. 

\paragraph{Data signatures and data alphabets.}
A \emph{data signature} $\dsig$ is a set of pairs 
$\{ \field_i : \mathit{type}_i \}_{i\in[n]}$, where 
$\field_i$ is a field name, and 
$\mathit{type}_{i}$ is the type of $\field_i$.
Common types include integers, 
floating point rationals and real numbers, 
the Boolean type $\bool$ and the bit vectors of fixed length.
An {\em evaluation} $\nu$ of a data signature 
$\dsig$ is a map that associates each field name $\field$ in $\dsig$ with a value of the corresponding type, denoted by $\nu.\field$. 
We denote the set of all evaluations of $\dsig$ by $\langE(\dsig)$. Henceforth, we will use the term {\em symbol} to refer to an element in $\langE(\dsig)$, while we refer to the set $\langE(\dsig)$ as the {\em data alphabet} of $\dsig$.  


\paragraph{Data languages.} A {\em data word} over a data signature $\dsig$ 
is 
a finite sequence $w = w_1 w_2\ldots w_n$ where each element $w_i$ is a symbol of $\langE(\dsig)$. 
The {\em length} of a data word $w$, denoted by $|w|$, is the number of symbols in the sequence. The {\em empty data word}, denoted by $\epsilon$, is the data word with no data symbols, i.e., $|\epsilon|=0$.
We denote the set of all data words over $\langE(\dsig)$ by $\langE(\dsig)^*$. 
A {\em data language} over $\langE(\dsig)$ is any subset of $\langE(\dsig)^*$. 
For any data word $w$ of length $n$ and $i \in [n]$, we denote its \emph{prefix} $w_1 \ldots w_i$ by $w_{\leq i}$,  and its \emph{suffix} $w_i \ldots w_n$ by $w_{\geq i}$.


\paragraph{Data logic.}  
This paper extends {\LTLf} by integrating data constraints expressed in first-order logic with equality, following standard syntax and semantics~\cite{DBLP:conf/unu/MannaZ02}. To handle multiple data types in the fields of symbols forming data words, we utilize many-sorted signatures. Specifically, we employ a many-sorted first-order logic $\dtheory$ with sorts $\mathit{data}_1,\ldots,\mathit{data}_n$. Each $\mathit{data}_i$ has a corresponding logic $\dtheory_{\mathit{data}_i}$, including function symbols of type $\mathit{data}_i^h\rightarrow \mathit{data}_i$ and relation symbols of type $\mathit{data}_i^h\rightarrow\bool$, with arity $h$. These logics encompass features like integer or real arithmetic, arrays, etc.
Henceforth, we assume that $\dtheory$ comprises constant symbols $\theoryconsts$, relation symbols $\theorypreds$, and function symbols $\theoryfuns$.


\section{{\LTLf} Modulo Theories}\label{sec: def LTLfMT}

We present {\LTLMT}, an extension of $\LTLf$~\cite{DBLP:conf/ijcai/GiacomoV13}, tailored for expressing data-word properties via embedded $\dtheory$-formulas. 
\LTLMT\ is a fragment of Geatti et al.'s logic~\cite{DBLP:conf/ijcai/GeattiGG22}, distinguished by the absence of quantified variables,
and it extends the logic of Felli et al.~\cite{DBLP:conf/aaai/FelliMPW23} by allowing more general data constraints. Henceforth, we assume that $\dsig$ is the data signature of the data words under consideration.

\paragraph{Syntax.}
{\bf Terms} of $\LTLMT$ are defined as follows:
$$
t \coloneqq \Next^n \field \mid {\WNext}^n \field \mid c \mid f(t_1,\ldots,t_k) \,,
$$
where $n\geq 0$, $\field$ is a field in
$\dsig$, $c\in\theoryconsts$, and $f$ is a $k$-arity function symbol in $\theoryfuns$. The grammar includes the {\em next term constructor} $\Next^n$ and the {\em weak next term constructor} ${\WNext}^n$. The superscript $n$ denotes ``$n$ steps in the future'' (called \emph{lookahead}).
For instance, $\Next^2 x$ represents the value of field $x$ two steps ahead, requiring those steps to exist. ${\WNext}^2x$ denotes the same value but does not require those steps to exist. 
Instead of $\Next^0 \field$ and ${\WNext}^0\field$, we can simply write $\field$.


\paragraph{Formulas} \hspace{-0,35cm} of $\LTLMT$ are generated by the following
grammar:
$$
    \phi \coloneqq \rel(t_1,\ldots,t_k) \mid \neg \phi \mid \phi_1 \wedge \phi_2 \mid
    \phi_1 \Until \phi_2 \mid \Tom \phi \,,
$$
where $\rel \in \theorypreds$
is a relation (or \emph{predicate}) symbol 
of arity $k$ applied to terms $t_1, \ldots, t_k$. 
We use $\neg$ for negation, $\wedge$ for conjunction, and $\Until$ for the {\em until operator}. The temporal operator $\Tom$ signifies {\em tomorrow} (distinct from {\em next} for clarity). 
In the grammars, $f(t_1,\ldots,t_k)$ and $\rel(t_1,\ldots,t_k)$ are assumed to be well-typed, ensuring that the function $f$ and relation $\rel$ can operate on the data types of $t_1,\ldots,t_k$.

\paragraph{Semantics.}
The $\LTLMT$ formulas are interpreted over data words, in accordance with an interpretation $\mathbb{D}$ of the symbols of the data logic $\dtheory$. 
For a given data word $w$, a term $t$ within the formula is interpreted as a value $\sem{t}_w$ derived from the underlying data logic. However, if the term attempts to read beyond the end of the word, it will be assigned one of two special values: $\err$ (a strong error) or $\Werr$ (a weak error). 
In particular, the semantics for function terms gives higher priority
to strong errors compared to weak errors arising from the arguments. Formally:
\begin{align*}
    \sem{\Next^n \field}_w &= \begin{cases}
    w_{n+1}.\field &\text{if } |w| \geq n+1 \\
    \err &\text{otherwise}.
    \end{cases} \\
    \sem{{\WNext}^n\field}_w &= \begin{cases}
    w_{n+1}.\field &\text{if } |w| \geq n+1 \\
    \Werr &\text{otherwise}.
    \end{cases} \\
    \sem{c}_w &= \mathbb{D}(c) \\
    \sem{f(t_1,\ldots,t_k)}_w &= \begin{cases}
        \mathbb{D}(f)(\sem{t_1}_w, \ldots, \sem{t_k}_w) \\
        \qquad\text{if }
        \sem{t_i}_w \not\in \{ \err, \Werr \} \text{ for all } i \\
        \err \quad\text{if } \sem{t_i}_w = \err \text{ for at least one } i \\
        \Werr \quad\text{otherwise}.
    \end{cases}
\end{align*}
Formulas of $\LTLMT$ are interpreted over data words. For a data word $w$ and a $\LTLMT$ formula $\phi$, we define the {\em satisfaction relation} $w\models \phi$ (i.e., $w$ is a model of $\phi$) as follows:
\begin{align*}
  w &\models \rel(t_1,\ldots,t_k) &\text{ iff } 
    &\Big( \sem{t_i}_w \not\in \{ \err, \Werr \} \text{ for all } i, 
    \text{ and } \\
    &&&\,\,\,\,\,\mathbb{D}(p)(\sem{t_1}_w, \ldots, \sem{t_k}_w) \text{ holds} \Big) \text{ or } \\
    &&&\Big( \sem{t_i}_w \neq \err \text{ for all } i, \text{ and }\\
    &&&\quad\sem{t_i}_w  = \Werr \text{ for at least one } i \Big) \\
  w &\models \neg\phi &\text{ iff } &w \not\models \phi \\
  w &\models \phi_1 \wedge \phi_2 &\text{ iff } &w \models \phi_1
  \text{ and } w \models \phi_2 \\
  w &\models \phi_1 \Until \phi_2 &\text{ iff } &\text{there is }
  i \in [|w|] \text{ s.t.\ } w_{\geq i} \models \phi_2, \text{ and } \\
  &&&\text{for all } j \in [i-1] \text{ it holds } w_{\geq j} \models \phi_1 \\ 
  w &\models \Tom \phi &\text{ iff } &|w|>1 \text{ and } w_{\geq 2} \models \phi.
\end{align*}

The semantics for predicates gives higher priority to strong errors
over weak errors: if any of the terms within the predicate yields a strong error, the predicate is false; otherwise, if weak errors arise from the terms, the predicate is true. The predicate is
evaluated according to its meaning in the data theory
only when no errors occur.
We define the \emph{language} of an $\LTLMT$ formula $\varphi$, denoted by $\langW(\varphi)$, as the set of all data words $w\in \langE(\dsig)^*$ such that $w\models \phi$.

\paragraph{Derived operators.} 
We also introduce supplementary logical and temporal operators derived from the fundamental operators outlined in the grammar for defining formulas. The logical operators include $\vee$ for disjunction, 
$\rightarrow$ for implication, $\leftrightarrow$ for bi-implication, and the constants $\true$ and $\false$. The additional temporal operators are $F\phi$ (eventually) and $G\phi$ (globally), representing $\mathit{true}\,\Until \phi$ 
and 
$\neg F\neg \phi$, 
respectively.
Using the standard Tomorrow operator $\Tom$, we define the {\em weak Tomorrow operator} $\WTom$ as 
$\WTom\phi \defeq \mathit{last} \vee \Tom\phi$, where $\mathit{last}$ is the formula $\neg \Tom\true$.
While $\Tom$ requires the existence of the next symbol in the word, $\WTom$ does not. Therefore, $\WTom \phi$ is always true on the last symbol of a data word.
We write $\Tom^n$ and $\WTom^n$ to signify $n$ repetitions of
the corresponding operator.

\begin{example} \label{ex:temperature}
Consider a simple {\bf temperature control system} 
with four variables:  a Boolean  $\mathit{heat}$ indicating heating status, a real-valued  $\mathit{temp}$ that represents the temperature in the building, and two integer-valued variables $e$ (energy consumption) and $t$ (time). The system follows these rules:
\begin{itemize}
  \setlength\itemsep{0.0em}
    \item Time is measured in discrete units of hours.

    \item Heating increases temperature by 1.5 degrees per hour.

    \item Without heating, the temperature decreases by 1 degree hourly. 

    \item Minimum heating duration is 4 hours.

    \item Minimum heating-off duration is 2 hours.

    \item Each heating hour consumes 1 unit of energy.

    
    
    
    
\end{itemize}
These rules can be expressed with the {\LTLMT} formula $\phi_\mathit{rules}$:
\begin{align*}
 &G \Big( \quad\,{\WNext}\,t = t+1\\
&\quad\,\,\wedge\,
\mathit{heat} \rightarrow ({\WNext}\,e = e+1 \,\wedge\, {\WNext}\,\temp = \temp +1.5) \\
&\quad\,\,\wedge\neg\mathit{heat} \rightarrow ({\WNext}\,e = e \,\wedge\, 
{\WNext}\,\temp = \temp - 1 ) \\
&\quad\,\,\wedge (\neg\mathit{heat} \wedge \Tom \mathit{heat}) \rightarrow (
\WTom^2 
\mathit{heat} \wedge 
\WTom^3 
\mathit{heat} \wedge 
\WTom^4 
\mathit{heat} ) \\
&\quad\,\,\wedge(\mathit{heat} \wedge \Tom \neg\mathit{heat}) \rightarrow \WTom^2 
\neg\mathit{heat}
\qquad\qquad\qquad\qquad
\,\,\,\,\Big)\\
&\wedge e=0 \wedge t=0.
\end{align*}
\end{example}

The {\bf satisfiability problem} (SAT) for a given {\LTLMT} formula $\phi$ with an associated data signature $\dsig$ asks whether $\langW(\phi)$ is empty.
The following theorem can be proved by reducing the halting problem for 2-counter machines~\cite{Minsky} to it (see the Appendix).

\begin{theorem}\label{thm: undecidability SAT problem} The SAT problem for {\LTLMT} is undecidable.
\end{theorem}

\paragraph{Normal Form.}
We show that removing the weak next term constructor ${\WNext}\,$ from the syntax does not affect expressiveness. 
This result is used to streamline the automaton construction outlined later.

\begin{lemma} \label{lem:no-weak}
For all $\LTLMT$ formulas $\phi$, there is an equivalent $\LTLMT$ formula
$\phi'$ without the weak next term constructor ${\WNext}\,$.
\end{lemma}
\begin{proof}
Consider a subformula $\theta$ of $\phi$ of the form $p(t_1,\ldots,t_k)$, where at least one term includes the weak next term constructor ${\WNext}\,$. Let $a$ (resp., $b$) denote the maximum lookahead of ${\WNext}\,$ (resp., $\Next$) in $t_1,\ldots,t_k$. For each $i\in[k]$, let $t'_i$ be the same as $t_i$ except that ${\WNext}^n$ is replaced by $\Next^n$. Replace $\theta$ with
$
\big( p(t'_1, \ldots, t'_k) \vee \neg\Tom^a \true \big) \wedge \Tom^b \true \,.
$
The formula $\phi'$ is obtained by applying the above substitution to all
subformulas of the type $p(t_1,\ldots,t_k)$ occurring in $\phi$.
It is straightforward to prove the equivalence of $\phi$ and $\phi'$.
\end{proof}

Then, we demonstrate that the lookahead can be limited to the values $\{0, 1\}$ while maintaining (un)satisfiability, at the cost of enlarging the signature, similarly to Lemma 7 in \cite{DBLP:conf/aaai/FelliMPW23}. We do not use this result in the automaton
construction in Section~\ref{sec:Automata Theoretic Construction},
because, while the enlarged signature is inconsequential for satisfiability checking, it makes model-checking and monitoring applications more cumbersome, and we want the automaton to work uniformly on all three applications.

\begin{lemma} \label{lem:no-lookahead}
For all $\LTLMT$ formulas $\phi$ without ${\WNext}\,$,
there exists an equi-satisfiable $\LTLMT$ formula
$\phi'$ in which every term of the form $\Next^n$ satisfies $n \in \{0, 1\}$.
\end{lemma}
\begin{proof}
Let $\dsig$ be the signature of $\phi$, and let $n^*$ be the maximum lookahead
of $\Next$ in $\phi$.
To decrease the lookahead to a maximum of 1, 
we enrich the signature with $n^*$ extra copies of the data fields,
i.e., we set:
$$\dsig' \defeq \bigcup_{i=0}^{n^*} \big\{ (\field_i : \mathit{type}) \mid
(\field : \mathit{type}) \in \dsig \big\} \,.$$
Then, let $\psi$ be the formula on $\dsig'$ obtained from $\phi$ be replacing 
each term of the type $\Next^n \field$ with $\field_n$.
The formula $\phi'$ required by the statement is the
following:
$$
\phi' \,\defeq\, \psi \wedge G \bigwedge_{i=1}^{n^*}\left( \Next^1 \field_{i} = \field_{i-1}\right)\,.
$$
Proving that $\phi$ and $\phi'$ are equisatisfiable is straightforward.
\end{proof}

\section{Symbolic  Data-Word Automata}\label{sec:Data Word Automata}
In this section, we introduce {\em Symbolic Data-Word Automata} ({\SDWA}s), a new automaton type extending traditional finite state automata (FA) to work with data words. Unlike FAs, {\SDWA}s utilize two distinct data signatures to define alphabets and state sets. The transition function includes constraints from a given data logic $\dtheory$.
{\SDWA}s can be seen as a linear (as opposed to branching) variant of \emph{symbolic data-tree automata} from~\cite{DBLP:conf/cav/FaellaP22}.
In the next section, we use {\SDWA}s as the pivotal model allowing us to address various decision problems related to {\LTLMT}, by reducing them to the emptiness problem for {\SDWA}s.

\begin{definition}
A {\bf symbolic data-word automaton} $\aut$ with data logic $\dtheory$, is a 5-tuple $(\dsig^\Sigma,\dsig^Q, \psi^{0}, \psi^\Delta, \psi^F)$ where: 
\begin{description}
    \item[$\dsig^\Sigma$] is the {\em alphabet data signature} defining the word alphabet $\Sigma = \langE(\dsig^\Sigma)$.
    Each field in $\dsig^\Sigma$ is typed with
    a sort of $\dtheory$;

    \item [$\dsig^{Q}$] is the {\em state data signature} defining the set of states 
    $Q = \langE(\dsig^Q)$.
    Each field in $\dsig^Q$ is typed with a sort of $\dtheory$;
    
    \item [$\psi^0(q)$] is a $\dtheory$-formula on the free variable $q$ of type $\dsig^Q$,\footnote{Since $q$ is a 
    variable of type  $\dsig^Q = \{ \field_i : \mathit{type}_i \}_{i\in[n]}$,  $\psi^0(q)$ becomes a $\dtheory$-formula on the variables $q.\field_1, \ldots, q.\field_n$.} defining the set of \emph{initial states}, 
    i.e., the set of all the elements $q\in Q$ such that $\psi^0(q)$ evaluates to true;
    
    \item[$\psi^\Delta(q, a, q')$] is the \emph{transition constraint}, defined as a $\dtheory$ formula,
    where $q$ and $q'$ are variables of type $\dsig^Q$, 
    and $a$ is a variable of type $\dsig^\Sigma$;
     
    \item[$\psi^F(q)$] is a $\dtheory$ formula on the free variable $q$, defining the set of \emph{final states} $F\subseteq Q$, i.e., the set of all the elements $q\in Q$ such that $\psi^F(q)$ evaluates to true.    
\end{description}

$\aut$ is designed to accept $\dsig^\Sigma$-words.
A data word $w$ is considered {\em accepted} by $\aut$ if there exists a function $\run: [0, l] \rightarrow Q$, where $l=|w|$, such that the following conditions hold:
(i) $\psi^0\big( \run(0) \big)$ is true, (ii) for all $i \in [0,l-1]$, $\psi^{\Delta}\big( \run(i), w_{i+1}, \run(i+1) \big)$ is true, and (iii) $\psi^F\big( \run(l) \big)$ holds.
\end{definition}

The set of all $\dsig^\Sigma$-words accepted by $\aut$ forms its {\em language}, denoted by $\langW({\aut})$.
    $\aut$ is {\bf deterministic} if
    \emph{(i)} for all 
    $q \in Q$ and 
    $a \in \Sigma$ there is exactly one state
    $q' \in Q$ such that $\psi^\Delta(q,a,q')$ holds,
    and \emph{(ii)} there is exactly one state $q$ such that
    $\psi^0(q)$ holds.
    The following result is proved in the Appendix
    and used to address the model-checking problem
    in Section~\ref{sec:model-checking}.

\begin{theorem} [{\sc Closure under Intersection}]\label{thm:closure_intersection}
Given two {\SDWA}s $\aut_1$ and $\aut_2$, we can effectively construct an {\SDWA} $\aut_{\cap}$ such that $\langW(\aut_{\cap})=\langW(\aut_1)\cap \langW(\aut_2)$.
\end{theorem}

\subsection{Solving the Emptiness Problem with {\CHC}s 
} \label{sec:SDWA-emptiness}
The {\em emptiness problem} for symbolic data-word automata consists in determining if a given {\SDWA} $\aut$ recognizes any word, i.e., whether $\langW(\aut)$ is empty. 

The undecidability of the problem follows from Theorem~\ref{thm: undecidability SAT problem} and the {\SDWA} construction 
in Section~\ref{sec:Automata Theoretic Construction}. However, we show that the
emptiness problem can be effectively reduced to the satisfiability of a {\CHC} system,
which are often solvable with efficient off-the-shelf solvers.




\paragraph{Constrained Horn Clauses.}
%
%
We fix a set $R$ of uninterpreted fixed-arity relation symbols representing the unknowns in the system. A {\em Constrained Horn Clause}, or {\CHC}, 
is a formula of the form 
$H\leftarrow C\wedge B_1\wedge\cdots\wedge B_n$
where:
\begin{itemize}
    \item $C$ is a constraint over the background data logic $\dtheory$ and does not include any application of 
    symbols in $R$;
    \item 
    $B_i$ is an application $\rel(v_1,\ldots,v_k)$ of a relation symbol $\rel\in R$ to first-order variables $v_1,\ldots,v_k$; 
    \item $H$ is the clause {\em head} and, similarly to $B_i$, is an application $\rel(v_1,\ldots,v_k)$ of a relation symbol $\rel\in R$ to the first-order variables, or it is $\mathit{false}$;
    \item all first-order variables in the signature of predicates and constraints are implicitly universally quantified.
\end{itemize}

A finite set $\mathcal{H}$ of {\CHC}s is a \emph{system}, corresponding to the first-order formula obtained by taking the conjunction of all its {\CHC}s. 
The semantics of constraints is assumed to be given as a structure. A system $\mathcal{H}$ with relation symbols $R$ is \emph{satisfiable} if there exists an interpretation for each predicate in $R$ such that all clauses in $\mathcal{H}$ are valid under that interpretation. 

In constraint logic programming, every {\CHC} system $\mathcal{H}$ has a unique minimal model, computed as the fixed-point of an operator derived from its clauses~\cite{DBLP:journals/jacm/EmdenK76,DBLP:journals/jlp/JaffarM94}. This 
semantics justifies the correctness of the reduction defined below.



\paragraph{Reduction.}
We propose a linear-time reduction from the emptiness problem for {\SDWA}s to the satisfiability of {\CHC} systems. 
The translation of {\SDWA}s, akin to programs with scalar variables, into {\CHC}s exhibits similarities with the use of proof rules in software verification~\cite{DBLP:conf/pldi/TorreMP09,DBLP:conf/pldi/GrebenshchikovLPR12}.

Given an {\SDWA} $\aut=(\dsig^\Sigma,\dsig^Q, \psi^{0}, \psi^\Delta, \psi^F)$, with structured variables $q$ and $q'$ of type $\dsig^Q$, $a$ of type $\dsig^\Sigma$, and $h(\cdot)$ as an uninterpreted predicate, we map $\aut$ to the {\CHC} system $\mathcal{H}_{\aut}$, consisting of the following {\CHC}s:
\begin{equation*} 
\mathcal{H}_{\aut}\defeq\begin{cases}
h(q) &\leftarrow\,\psi^0(q) \\
h(q') &\leftarrow\, h(q) \wedge \psi^\Delta(q,a,q')\\
\mathit{false} &\leftarrow\, h(q) \wedge\psi^F(q).
\end{cases}
\end{equation*}


\begin{theorem}[{\sc Emptiness}]\label{thm:emptiness}
The language $\langW(\aut)$ of an {\SDWA} $\aut$ is empty iff 
the {\CHC} system $\mathcal{H}_{\aut}$ is satisfiable.
\end{theorem}

\section{The Automata-Theoretic Construction}\label{sec:Automata Theoretic Construction}

This section constructs a deterministic \SDWA equivalent to the provided \LTLMT\ formula $\varphi$ in three steps:
\begin{enumerate}
\item Convert $\varphi$ into an $\LTLf$ formula $\varphi'$ by abstracting away the data constraints.
\item Employ classical results to obtain a deterministic finite
automaton $\aut_{\varphi'}$ equivalent to $\varphi'$.
\item Convert $\aut_{\varphi'}$ into an \SDWA by checking the
data constraints on the appropriate data.
\end{enumerate}
The process is similar to a result for a fragment of the logic \MSOD~\cite{DBLP:conf/cav/FaellaP22}. In contrast, our paper presents a deterministic automaton and manages an arbitrary lookahead of the next term constructor. 


\subsection{Transforming to \LTLf}\label{sec:encoding}

Here, we define the $\LTLf$ formula $\toltl(\phi)$, crafted to abstract data constraints from $\phi$. 
Thanks to Lemma~\ref{lem:no-weak},
we can assume that $\phi$ does not contain
the ${\WNext}\,$ constructor. 

Consider all subformulas of the form $p(t_1,\ldots,t_k)$ within $\phi$, denoted as $\theta_1, \ldots, \theta_m$. For each $i\in[m]$, introduce a new atomic proposition $\prop_i$, and let $n_i$ be the maximum lookahead of the $\Next$ operator within $\theta_i$. Construct the $\LTLf$ formula $\toltl(\phi)$ by replacing each occurrence of $\theta_i$ in $\phi$ with the formula $\Tom^{n_i} \prop_i$. 
The rationale behind this transformation is that to establish the truth of $\theta_i$, it may be necessary to read from the data word up to $n_i$ positions ahead. As our ultimate goal is to construct a deterministic FA for $\phi$, we defer the verification of $\theta_i$ by $n_i$ steps, allowing the automaton to read and store all relevant data needed to establish its truth. 
Let $\AP = \{ \prop_1, \ldots, \prop_m \}$ be the atomic propositions in $\toltl(\phi)$.

To establish the semantic relation between $\phi$ and $\toltl(\phi)$, we apply a similar abstraction to data words. Intuitively, we need to define the truth of each atomic
proposition $\prop_i$ based on the data present in the data word.
To preserve the correspondence between $\Tom^{n_i} \prop_i$
and $\theta_i$,
the truth of $\prop_i$ in a given position $j$ in the abstract word must be the same as the value of $\theta_i$ in the data word
$n_i$ steps earlier than $j$.
It is still not clear what should be the value of $\prop_i$
in the first $n_i$ positions of the abstract word, because 
for those positions there is not enough past data to evaluate
$\theta_i$.
To solve this technical issue, we assume that each data type includes
a \emph{default value}, and we use those default values to
evaluate $\theta_i$ when it reads beyond the start of the data word. This solution ends up attributing essentially arbitrary values to $\prop_i$ in the early positions of the abstract word,
but this does not thwart our construction.
Indeed, since every occurrence of $\prop_i$ in $\toltl(\phi)$
appears in the context $\Tom^{n_i} \prop_i$,
$\toltl(\phi)$ is insensitive to the value of $\prop_i$
in the first $n_i$ positions of the word.

Let $\mathit{def}_k$ denote the data word of length
$k$ where all fields hold the default value of their type.
For a data word $w$ on the
data signature $\dsig$ of $\phi$,
let $\toltl(\phi, w)$ be the word $w'$ over the finite alphabet $2^\AP$, defined as follows:
\begin{itemize}
    \item the length of $w'$ is the same as the length of $w$;
    \item for all $i=1,\ldots,m$ and 
    $j = 1, \ldots, |w|$,
    $$
    \prop_i \in w'_j \iff \begin{cases}
        w_{\geq j - n_i} \models \theta_i
    &\text{if } j > n_i \\
        \mathit{def}_{n_i - j+1} \cdot w \models \theta_i
    &\text{otherwise.}
    \end{cases}        
    $$   
\end{itemize}

\begin{theorem} \label{thm:abs-word}
For all $\LTLMT$ formulas $\phi$ and data words $w$
on the same data signature,
$w\models\phi$ iff $\toltl(\phi,w)\models\toltl(\phi)$.
\end{theorem}

\subsection{Building the \SDWA} \label{sec:sdwa}

Recall the following well-known result about $\LTLf$:
\begin{theorem}[\cite{DBLP:conf/aips/GiacomoF21}]
For all $\LTLf$ formulas $\phi$,
we can effectively construct an equivalent deterministic FA.
\end{theorem}

Let $\aut_{\varphi'} = (2^\AP, Q, { q_0 }, \delta, F)$ be a deterministic FA equivalent to $\varphi' = \toltl(\varphi)$, where 
$\delta: Q \times 2^\AP \to Q$.
%
We define a deterministic \SDWA $\aut_\varphi = (\dsig, \dsig^Q, \psi^0, \psi^\Delta, \psi^F)$ with:
\begin{description}
\item[$\dsig^Q$] $= \{ \mathit{state} : Q \} \cup \{ \mathit{data}^i : \dsig \}_{i \in [N]} \,,$ 
where $N$ is the maximum of the $n_i$'s, that is,
the maximum lookahead of the $\Next$ term constructor
in $\phi$.
In words, the state data signature of the
automaton holds a state of the finite automaton $\aut_{\varphi'}$
and $N$ copies of the data signature of the original formula.
The first component (i.e., $\mathit{state}$) is used to simulate a run of $\aut_{\varphi'}$, and the second 
(i.e., $\mathit{data}$) stores the previous $N$ symbols read from the data word.
Specifically, the field $\mathit{data}^i$ contains the symbol
that was read $i$ steps ago, as specified in the definition
of $\psi^\Delta$ below.
\item[$\psi^0(s)$] holds iff $s.\mathit{state} = q_0$,
and all other fields contain the default value of their type.
\item[$\psi^\Delta(s, a, s')$] holds iff 
\emph{(i)} $s'.\mathit{data}^1 = a$, 
\emph{(ii)} $s'.\mathit{data}^{i+1} = s.\mathit{data}^{i}$ for all $i \in [N-1]$, and
\emph{(iii)} it holds $\delta(s.\mathit{state}, \sigma) = s'.\mathit{state}$, where, for all $i\in [m]$:
\begin{multline} \label{symbolic-delta}
\prop_i \in \sigma \text{ iff }
s.\mathit{data}^{n_i} s.\mathit{data}^{n_i-1} \cdots s.\mathit{data}^{1}  a \models \theta_i \,.
\end{multline}

\item[$\psi^F(s)$] holds iff $s.\mathit{state} \in F$.
\end{description}

Note that when the symbolic automaton begins reading a data word, its \emph{buffer} of symbols kept in the $\mathit{data}$ fields is initialized with default values, which are gradually replaced by those contained
in the last $N$ symbols read.
Hence, the data constraints $\theta_i$ are initially evaluated
over default data values, resulting in essentially arbitrary truth values.
This is not a problem, because the formula $\varphi'$, which guides the 
$\mathit{state}$ component of the symbolic automaton and prescribes which data constraints must be true
at each step, is insensitive to the value of $\prop_i$
in the first $n_i$ positions of the word, as noted earlier.



We now state the main result of this section.
\begin{theorem} \label{thm:sdta2}
For all \LTLMT\ formulas $\varphi$, we can effectively construct a deterministic {\SDWA} $\aut_\varphi$ s.t. $\langW(\aut_\varphi) = \langW(\varphi)$. 
\end{theorem}

\section{Applications}

In this section, we demonstrate the versatility of our automata-theoretic approach in addressing diverse challenges associated with {\LTLMT}, namely satisfiability, model checking, and monitoring.



\subsection{Satisfiability}

The findings in the preceding sections offer a direct route to tackle the satisfiability problem for {\LTLMT}.
Theorem~\ref{thm:sdta2} establishes that any {\LTLMT} formula can be converted into an equivalent \SDWA. By leveraging Theorem~\ref{thm:emptiness}, we transform the emptiness of an \SDWA into the satisfiability of a system of {\CHC}s. This reduction facilitates the application of advanced solution techniques, as illustrated in Section~\ref{sec:experiments}. There, we assess the satisfiability of diverse benchmark formulas using Z3 as the underlying {\CHC} solver.
\begin{theorem}
    The {\LTLMT} satisfiability problem can be effectively reduced to the emptiness problem for {\SDWA}s.
\end{theorem}


\subsection{Model Checking} \label{sec:model-checking}
The {\em {\LTLMT} model-checking problem} for an {\SDWA} $\mathcal{M}$ and an {\LTLMT} formula $\varphi$ is to decide whether there exists a data word $w$ in the language of $\mathcal{M}$ that is a model of $\varphi$ (i.e., $w\models \varphi$).

The undecidability of this problem stems from the undecidability of the emptiness problem for {\SDWA}s. Nonetheless, we can readily devise a sound procedure for solving the model-checking problem by ultimately employing {\CHC} solvers.

Our procedure is as follows. Firstly, we create an {\SDWA} $\aut_\varphi$ such that $\langW(\varphi) = \langW(\aut_\varphi)$ (refer to Theorem~\ref{thm:sdta2}). Subsequently, utilizing the construction associated with the closure of {\SDWA}s under intersection (as detailed in the proof of Theorem~\ref{thm:closure_intersection}), we generate a new {\SDWA} $\aut$ such that $\langW(\aut) = \langW(\mathcal{M}) \cap \langW(\aut_\varphi)$. The determination of the answer to the model-checking problem relies on establishing whether $\langW(\aut)$ is empty: a positive answer to the model-checking problem for $\mathcal{M}$ and $\varphi$ occurs iff the language of $\aut$ is non-empty.
 
\begin{theorem}
    The {\LTLMT} model-checking problem can be effectively reduced  to the emptiness problem for {\SDWA}s.
\end{theorem}

When the {\CHC} engine 
proves that the model-checking problem admits a positive answer, a data word satisfying the formula can be derived from the counterexample to the {\CHC} system.




\subsection{Runtime Monitoring}

At times, the model checking of complex systems becomes intractable owing to the inherent complexity of the system. This leads to the adoption of less intricate verification approaches, such as runtime monitoring. In runtime monitoring, we continuously observe the system's execution and promptly notify any violation or fulfillment of the specified property. In this context, we delve into runtime monitoring concerning a given {\LTLMT} formula $\varphi$.

In defining the monitoring problem, we trace the satisfaction status of $\varphi$ along a trace of $\mathcal{M}$, considering the trace fragment $w$ that has been observed thus far.
Following established conventions~\cite{DBLP:journals/logcom/BauerLS10,DBLP:conf/bpm/MaggiMWA11,DBLP:conf/aaai/FelliMPW23}, we define $\RV = \{\PS, \CS, \CV, \PV\}$ as the set of four \emph{monitoring states}: current satisfaction (\CS), permanent satisfaction (\PS), current violation (\CV), and permanent violation (\PV).
\begin{definition}
For a given data word (trace) $w$ and an {\LTLMT} formula $\varphi$, the {\bf monitoring problem} is the task of determining the monitoring state $\mst\in\RV$ that satisfies one of the following conditions:
 \begin{itemize}
    \item $\mst=\CS$, $w \models \varphi$, and $w w'\not\models\varphi$ for some trace $w'$;
    \item $\mst=\PS$, $w \models\varphi$, and $w w'\models\varphi$ for every trace $w'$; 
    \item $\mst=\CV$, $w \not\models \varphi$, and $w w'\models \varphi$ for some trace $w'$;
    \item $\mst=\PV$, $w \not\models \varphi$, and $w w' \not\models\varphi$ for every trace $w'$.\qed
\end{itemize}
\end{definition}

The following undecidability result should come as no surprise,
and can be proved along similar lines as the previous ones.
\begin{theorem} \label{thm:undecidable-monitor}
The monitoring problem for {\LTLMT} is undecidable (already for linear constraints).
\end{theorem}

We provide a sound procedure to solve 
the monitoring problem for {\LTLMT}. To achieve this, 
consider the 
deterministic \SDWA $\aut_\varphi = (\dsig, \dsig^Q, \psi^0,\allowbreak\psi^\Delta,\psi^F)$, as detailed in  Theorem~\ref{thm:sdta2}. Now, define two new {\SDWA}s as follows: 
\begin{description}
\item[$\aut_\varphi(q,F)$] is derived from $\aut_\varphi$ by replacing the initial state with the state $q\in \dsig^Q$, and maintaining the same final states as $\aut_\varphi$, i.e.,
$\psi^F(s)= \big( s.\mathit{state} \in F \big)$.
\item[$\aut_\varphi(q,\overline{F})$] is obtained from $\aut_\varphi$ by replacing the initial state with $q\in \dsig^Q$, and updating the final states with the formula
$\psi^F(s)= \big( s.\mathit{state} \not\in F \big)$.
\end{description}
\begin{mdframed}
\noindent{\bf Monitoring Procedure:} 
\begin{enumerate}
    \item Utilizing a structured variable, denoted as $q$, we monitor the state of $\aut_\varphi$ as the analyzed system evolves. Initialization of $q$ involves assigning it the unique state $s$ that satisfies $\psi^0(s)$.
    \item After each system transition, we update $q$ with the unique state in which $\aut_\varphi$ moves to. We determine the monitoring state as follows:
    \begin{description}
    \item[$\CS$:] if $\psi^F(q)=\true$, and $\langW\left(\aut_\varphi(q,\overline{F})\right)\not=\emptyset$; 
    \item[$\PS$:] if $\psi^F(q)=\true$, and $\langW\left(\aut_\varphi(q,\overline{F})\right)=\emptyset$; 
    \item[$\CV$:] if $\psi^F(q)=\false$, and $\langW\left(\aut_\varphi(q,F)\right)\not=\emptyset$; 
    \item[$\PV$:] if $\psi^F(q) = \false$, and $\langW\left(\aut_\varphi(q,F)\right)=\emptyset$. 
    \end{description}
\end{enumerate}
\end{mdframed}
The above procedure justifies the following result.
\begin{theorem}\label{thm: monitoring precise}
The \LTLMT\ monitoring problem can be effectively reduced to the
emptiness problem for {\SDWA}s.
\end{theorem}    

As per Theorem~\ref{thm:undecidable-monitor}, the procedure might time out. One approach is to execute $\aut_\varphi$ along with the system and provide a conclusive answer only upon the system's termination. 
However, approximate solutions that anticipate an answer remain viable. By performing two emptiness checks on FAs corresponding respectively to $\toltl(\varphi)$ and to its complement
(rather than on the {\SDWA}s $\aut_\varphi(q,\cdot)$), a correct answer is given when the result is $\PS$ or $\PV$. 
If the result is $\CS$ (resp., $\CV$), the actual monitoring state may be $\{ \CS, \PS\}$ (resp., one of $\{ \CV, \PV\}$).
This imprecision does not introduce monitoring errors but extends monitoring until a definitive answer is attainable.
    



\section{Satisfiability Experiments} \label{sec:experiments}

In this section, we present experiments  to evaluate the satisfiability of various \LTLMT\ formulas. The experiments include the temperature control system outlined in Example~\ref{ex:temperature} and formulas from \cite{DBLP:conf/ijcai/GeattiGG22}.


Figure~\ref{fig:toolchain}  illustrates the experimental toolchain.
We manually translated each \LTLMT\ formula into \LTLf, as described in Section~\ref{sec:encoding}, and then into \LTL, in order to feed it to the tool \href{https://spot.lre.epita.fr/}{SPOT}~\cite{DBLP:conf/cav/Duret-LutzRCRAS22}, generating a corresponding FA.
Next, our script converts the FA into a {\CHC} system encoding the emptiness of the corresponding symbolic {\SDWA}, as detailed in Sections~\ref{sec:sdwa} and~\ref{sec:SDWA-emptiness}, representing each FA state as a distinct uninterpreted predicate.
Finally, we use the SMT-solver Z3 to check the satisfiability of the {\CHC} system. Theorem~\ref{thm:emptiness} establishes an inverse link between the satisfiability of the original \LTLMT\ formula and the {\CHC} system. In this section, references to satisfiability refer to the original \LTLMT\ formula.
\tikzset{
  tool/.style={rectangle, align=center, draw = black, fill = white, thin, inner sep=4pt},
  doc/.style={rectangle, align=center, rounded corners=.2cm, fill = white, thin, inner sep=2pt},
  holds/.style = {-latex, line width=1.3pt, black!40, text=black}
}

\begin{figure}[t]
\centering
\begin{tikzpicture}[auto, node distance=1.2cm, shorten >= 1pt, shorten <= 2pt]
  \tikzstyle{every node}=[font=\scriptsize]
  \node[tool,minimum height=0.5cm] (A) at (6, -0.2) {{\CHC} solver (Z3)};
  \node[doc] (G) at (0, 0.5) {{\LTLMT}\\formula};
  \node[doc] (B) at (2.15, 0) {Data signature\\and constraints};
  \node[doc] (F) at (1.8, 1) {{\LTLf}\\formula};
  \node[tool, minimum height=0.5cm] (E) at (3.5, 1) {SPOT};
  \node at (6.5,0.5) {{\CHC}s};
  \node[tool,thick] (D) at (6, 1) {Automaton to {\CHC}s};
  \path[holds,shorten <= 0pt] (D) edge (A);
  \path[holds] (F) edge (E) (E) edge node[above] {FA} (D);
  \path[holds,shorten >= 8pt] (B) edge (D);
  \path[holds,dashed] (G) edge (F) (G) edge (B);
\end{tikzpicture}
\caption{Architecture of the prototype implementation.
Dashed transformations are performed manually. 
}
\label{fig:toolchain}
\vspace{1.0cm}
\end{figure}
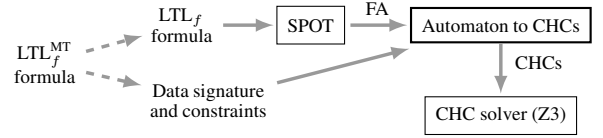

Experiments were performed on an AMD Ryzen 9 5900X (3.70Ghz) with 32GB of RAM and running Windows 10. The {\CHC} solver was Z3, version 4.6.0 (64 bit).
All formulas were also tested with the \href{https://github.com/black-sat/black}{BLACK tool}
by Geatti et al.~\cite{DBLP:conf/ijcai/GeattiGG22}.

The experimental results are presented in Table~\ref{tab:experiments}.

\paragraph{\bf Temperature Control.}
Reconsider Example~\ref{ex:temperature}, where the formula $\phi_\mathit{rules}$ describes 
a temperature control system. 
Suppose we want to verify the system's capability to bring the temperature back
to at least 20$^\circ$C after 24 hours, starting from 20$^\circ$C and never dropping below 18$^\circ$C,
with an energy budget of $N$ units, where $N$ is a parameter.
We express this requirement with the following 
formula: 
\begin{align*}
\tempctrl \defeq &\quad  \phi_\mathit{rules} 
    \wedge (\temp=20) \wedge G(\temp \geq 18) \\
    &
    \wedge X^{24} (e \leq N \,\wedge\, \temp \geq 20) \,.
\end{align*}
Verifying $\tempctrl$ satisfiability determines if the system can achieve the specified property. The resulting model provides an energy-efficient heating schedule. 
A binary search on $N$ showed that at least 10 hours of heating is needed. All cases were resolved within seconds, with BLACK being slightly faster.



\newcommand{\sat}{Yes}
\newcommand{\unsat}{No}

\begin{table}
\begin{center}
\small
\setlength{\tabcolsep}{4pt}
\begin{tabular}{lcrrrr}
&&&&\multicolumn{2}{c}{Time (sec.)}\\  \cline{5-6}
          Formula     &$N$ &Sat &\# of {\CHC}s &\makecell{This paper} &\makecell{BLACK \cite{DBLP:conf/ijcai/GeattiGG22}} \\ 
     \hline\hline
     \multirow{5}{*}{$\tempctrl$} 
               &6  &\unsat &\multirow{5}{*}{202}                                     &4.3 &$<1$ \\
               &9  &\unsat & &6.9 &$<1$ \\
               &10 &\sat   & &2.3 &$<1$ \\ 
               &12 &\sat   & &1.7 &$<1$ \\
               &24 &\sat   & &$<1$ &$<1$ \\     
               \hline
     \multirow{4}{*}{$\LIA_1$} 
               &-1 &\unsat &\multirow{4}{*}{7}  &$<1$ &timeout \\
               &10   &\sat & &$<1$ &$<1$ \\
               &100  &\sat & &$<1$ &$<1$\\
               &1000 &\sat & &6.6 &5.2 \\ \hline
     \multirow{3}{*}{$\LIA_2$} 
               &10  &\unsat &14 &$<1$ &$<1$ \\
               &50  &\unsat &54 &2.7 &4.8 \\
               &100 &\unsat &104 &59 &167.4 \\ \hline
     \multirow{3}{*}{$\LRA_1$} 
               &10   &\sat &18   &$<1$ &$<1$ \\
               &100  &\sat &108  &$<1$ &6.2 \\
               &1000 &\sat &1008 &6.9  &timeout\\ \hline
     $\imposs$  &- &\unsat &5 &$<1$ &timeout 
\end{tabular} 
\end{center}
\caption{Satisfiability experiments. Running times for this paper refer to the CHC solving time only, as the time to generate the {\CHC}s is negligible. Timeout is 10 minutes.
} \label{tab:experiments}
\vspace{0.3cm}
\end{table}

\paragraph{\bf Formulas from Geatti et al.}
We analyzed all formulas from Table~2 in~\cite{DBLP:conf/ijcai/GeattiGG22}, which includes two families
with constraints in linear integer arithmetic (LIA), denoted by $\LIA_1$ and $\LIA_2$, and a family with constraints in linear real arithmetic (LRA), denoted $\LRA_1$.
All formulas are parameterized by an integer $N$.
Table~\ref{tab:experiments} shows that our approach handles
all instances, showing similar or superior performance. 
Notably, we can detect the unsatisfiability
of $\LIA_1$ with parameter $-1$. 
Moreover, ~\cite{DBLP:conf/ijcai/GeattiGG22} contains two families of formulas that our approach cannot currently handle,
because the theories employed are not supported by {\CHC} solvers:
one with a non-linear real constraint;\footnote{In that paper,
that family is erroneously tagged with \emph{LRA}, despite containing the non-linear term
$\frac{1}{g}$.}
and another with an uninterpreted integer function.
Lastly, the authors of~\cite{DBLP:conf/ijcai/GeattiGG22} point out that the unsatisfiability of the formula
$$
\imposs \:\defeq\: G(x>3) \wedge F(x<2)$$
cannot be proved by their method, whereas
our approach promptly identifies the contradiction.


\section{Related Work}\label{sec:related work}
Our work is related to many works in the literature in different ways. In addition to the works discussed in the introduction, here we focus on those that seem to be closest to the results presented in this paper.

\paragraph{Data-aware Logics.}
Various proposals extend temporal logics with
data-oriented features.
Temporal Stream Logic \cite{DBLP:conf/cav/Finkbeiner0PS19} is an undecidable logic aimed at reactive synthesis, 
able to model the dynamic input-output relation 
of a reactive system while abstracting from specific data domains and operations.
Further developments support specific data theories \cite{finkbeiner2022temporal}, 
develop a counterexample-guided
abstraction refinement (CEGAR) loop involving {\LTL} synthesis and
SMT queries \cite{DBLP:conf/fmcad/MaderbacherB22},
and exploring connections with  syntax-guided synthesis \cite{DBLP:conf/pldi/ChoiFPS22}.

Other {\LTL} extensions aim at retaining the decidability of satisfiability by focusing on simple data theories \cite{DBLP:journals/iandc/DemriD07} or  by avoiding comparisons between data at different positions of the word
\cite{DBLP:conf/cav/DietschHLP15,baader2015temporal,DBLP:conf/aaai/GianolaMW24}.\footnote{Note that, in the context of model checking, 
comparing data in different positions can generally be achieved within the system, using extra storage.
Hence, a more detailed comparison with the concurrent findings of Gianola et al.~\cite{DBLP:conf/aaai/GianolaMW24}
merits further investigation.}
%
Additionally, Geatti et al.~\cite{DBLP:conf/ecai/GeattiGGW23}
describe various decidable fragments of $\LTLMT$.
In contrast, our logic (derived from \cite{DBLP:conf/ijcai/GeattiGG22})  is data-theory-agnostic and allows comparisons between data values at different positions using the next term constructor. 
These features radically raise
the expressive power of the specifications and renders the 
satisfiability problem undecidable.
Thus, the above extensions of \LTL\ do not
subsume our results, and our contributions do not subsume the
state-of-the-art decidability results.

\paragraph{Symbolic Automata.}


There is a variety of models for symbolic automata,
the most prominent being the symbolic finite automata of
D’Antoni and Veanes~\cite{d2021automata}. 
In these automata, only the alphabet is
possibly infinite, whereas the state space is finite. In particular, they do not allow to compare data values in different
positions of the data word, which is a main concern of our investigation. Such symbolic finite automata form a strict and
decidable subclass of our {\SDWA}s. 
In contrast, {\SDWA}s support infinite state spaces and their emptiness is undecidable, similarly to programs.

Several other models feature a hybrid state-space with a finite component that has a limited interaction with the infinite domain of symbols being read.
For instance, the Theory Mealy and Moore machines of Maderbacher and Bloem~\cite{DBLP:conf/fmcad/MaderbacherB22} pair a finite state-space with the possibility of storing data values and
comparing them with predicates from the given data theory.
Other models of finite automata on infinite alphabets
include the register and pebble machines of Neven et al.~\cite{neven2004finite} and the variable automata of
Grumberg et al.~\cite{grumberg2010variable}.

\paragraph{CHCs for Program Verification.}
Our work builds upon the recent surge of interest in constrained Horn clauses for automated software model checking~\cite{DBLP:conf/birthday/BjornerGMR15,DBLP:journals/corr/BeyenePR16}. Algorithms that efficiently solve systems of CHCs have been developed, often by adapting or extending techniques from automatic program verification~\cite{DBLP:conf/pldi/GrebenshchikovLPR12,DBLP:conf/synasc/GurfinkelB19,DBLP:conf/birthday/BjornerGMR15}. {\CHC}s offer a unique and elegant way to construct model checkers entirely through logic rules~\cite{DBLP:conf/aaai/Faella23,DBLP:journals/jar/ChampionCKS20,DBLP:journals/corr/GarocheKT16,DBLP:conf/cav/GurfinkelKKN15,DBLP:conf/fm/HojjatKGIKR12,DBLP:conf/cav/KahsaiRSS16,DBLP:conf/pldi/KobayashiSU11,DBLP:conf/esop/0002T020}.
This approach holds promise for efficient verification of temporal logic with data, especially considering the significant advancements made in {\CHC} satisfiability solvers in recent years~\cite{DBLP:journals/corr/abs-2109-04635}.

To leverage this potential, we introduce a novel automata-theoretic approach~\cite{r1994computer,DBLP:conf/lics/VardiW86}. We employ a translation technique that transforms {\SDWA}s (or equivalently, transition systems) into {\CHC}s. This translation aligns with existing applications of {\CHC}s in software verification~\cite{DBLP:conf/pldi/TorreMP09,DBLP:conf/pldi/GrebenshchikovLPR12}. Furthermore, our approach follows the growing trend of using automata theory in automated software model checking~\cite{DBLP:conf/cav/HeizmannHP13}. 

\paragraph{Runtime Monitors.} 
Related monitoring problems have been studied in~\cite{DBLP:journals/sttt/DeckerLT16}, focusing on comparing data values at arbitrary distances, and~\cite{DBLP:conf/aaai/FelliMPW23}, with an emphasis on linear arithmetic and properties conducive to decidable solutions.
Unlike~\cite{DBLP:conf/aaai/FelliMPW23}, our approach, while possibly non-terminating, operates on more general data constraints. While our method might not always find a solution, it can handle a wider range of data constraints compared to these prior works.

\section{Conclusions} \label{sec: conclusions}
In summary, we presented an novel automata-based approach supporting linear temporal logic modulo theory as a specification language for data words. Our framework translates {\LTLMT} into symbolic data-word automata, demonstrating their efficacy in capturing intricate temporal properties of executions of infinite-state systems.

Notwithstanding the undecidability of both {\LTLMT} satisfiability and {\SDWA} emptiness, we showed that both problems can be reduced to solving {\CHC}s, which allows us to capitalize on modern solvers. Empirical experiments substantiate the effectiveness of our approach, sometimes outperforming a previous custom solution and tool. These results highlight the practicality and broad applicability of our automata-based framework. 
Our approach seamlessly extends beyond satisfiability to encompass model checking and runtime monitoring in a unified way, effectively bridging three traditionally separate problems in prior research.


\paragraph{Future work.} 
A promising direction for future work is integrating results from
~\cite{DBLP:conf/aaai/FelliMPW23}and~\cite{DBLP:conf/ijcai/GeattiGG22},
into the domain of solving {\CHC}s,
to improve the effectiveness of {\CHC} solvers in general and our approach in particular.
%
%
Additionally, we see potential in extending our framework to accommodate other LTL dialects, such as CaRet logic~\cite{DBLP:conf/tacas/AlurEM04} with data and context-free systems that involve data. This expansion could advance formal verification and monitoring by handling a wider range of temporal specifications.

\smallskip

\noindent{\bf Funding Information.}
Supported by INDAM-GNCS 2023-24, 
the MUR project SOP (Securing sOftware Platforms - CUP: H73C22000890001) 
part of the SERICS project (n. PE00000014 - CUP: B43C22000750006), 
{\em Verifica di proprietà di sicurezza nello sviluppo del
software} under the Start-up 2022 program funded by the Computer Science Division
UNIMOL, and the MUR project Future AI Research (FAIR) Spoke 3.

\bibliography{ECAI}

\clearpage
\nobalance

\appendix

\section{Undecidability of the Satisfiability Problem}\label{app:undecidability}

The satisfiability problem for a given {\LTLMT} formula $\phi$ with an associated data signature $\dsig$ asks whether there exists a data word $w\in \langE(\dsig)^*$ such that $w\models \phi$. We can prove that the satisfiability problem for {\LTLMT} is undecidable by reducing to it the halting problem for 2-counter machines~\cite{Minsky}.
Let $\dtheory$ be the quantifier-free theory of linear integer arithmetic. To model the execution of any given 2-counter machine, we use data words whose signature has two fields of type $\mathbb{N}$ to represent the counters and an enumeration field to track the current instruction. Each machine configuration corresponds to a symbol in the word, and we can use constraints in {\LTLMT} to ensure that two consecutive symbols in the data word model a machine transition. We can also express the property of a halting computation in our logic. Thus, even though the underlying data logic $\dtheory$ is decidable, the satisfiability of {\LTLMT} is undecidable.
Decidability can be regained for $\LTLMT$ if we interpret the underlying data logic over a finite domain. In this case, the satisfiability problem becomes equivalent to that of the standard $\LTLf$.

\section{Closure of {\SDWA}s under Union and Intersection}

The set of data languages accepted by {\SDWA}s is effectively closed under union and intersection. That means that if $L_1$ and $L_2$ are two such languages, so are $L_1\cup L_2$ and $L_1\cap L_2$.
Theorem~\ref{thm:closure_intersection} is an immediate consequence of the
following result.

\begin{theorem}[{\sc Closure under Union and Intersection}]\label{th:closure_union_intersection}
Given two {\SDWA}s $\aut_1$ and $\aut_2$, we can effectively construct two {\SDWA}s $\aut_{\cap}$ and $\aut_{\cup}$ such that:
\begin{itemize}
    \item $\langW(\aut_{\cap})=\langW(\aut_1)\cap \langW(\aut_2)$, and
    \item $\langW(\aut_{\cup})= \langW(\aut_1)\cup \langW(\aut_2)$.
\end{itemize}
\end{theorem}
\begin{proof}
 
Let $\aut_i=(\dsig^\Sigma_i,\dsig^Q_i, \psi_i^0, \psi^\Delta_i, \psi^F_i)$, for $i\in\{1,2\}$. We assume, without loss of generality, that the field names $F_1$ of $\dsig^Q_1$ and the field names $F_2$ of  $\dsig^Q_2$ do not overlap, i.e., $F_1\cap F_2\not=\emptyset$. Also, let $\aut_\sigma=(\dsig^\Sigma_\sigma,\dsig^Q_\sigma, \psi_\sigma^0, \psi^\Delta_\sigma, \psi^F_\sigma)$, for $\sigma\in\{\cap,\cup\}$. 

We design $\aut_\cap$ and $\aut_\cup$ to simultaneously simulate $\aut_1$ and $\aut_2$  on the same input data word. The definition of $\aut_\sigma$ closely resembles the cross-product construction used in standard finite state automata. Therefore, the correctness proof follows a similar pattern. 

Formally, the construction is as follows. For a state $q$ of $\aut_\sigma$, we denote by $q_{|F_i}$ the record containing only the elements of  $q$ corresponding to the fields in $F_i$. For $\sigma\in\{\cap,\cup\}$, we define the components of $\aut_\sigma$ as follows:
\begin{description}
    \item[States:] $\dsig^Q_\cap\defeq\dsig^Q_\cup\defeq\dsig^Q_1\cup\dsig^Q_2$;
    \item[Initial states:] $\psi^0_\sigma(q)\defeq\psi^0_1(q_{|F_1})\wedge\psi^0_2(q_{|F_2})$;
    \item[Transition function:] $\psi^\Delta_\sigma(q,a,q')\defeq\psi^\Delta_1(q_{|F_1},a,q'_{|F_1})\wedge\psi^\Delta_2(q_{|F_2},a,q'_{|F_2})$;
    \item [Final states:]$\psi^F_\cap(q)\defeq\psi^F_1(q_{|F_1})\wedge\psi^F_2(q_{|F_2})$, and $\psi^F_\cup(q)\defeq\psi^F_1(q_{|F_1})\vee\psi^F_2(q_{|F_2})$.
\end{description}
\end{proof}

\section{Proof of Theorem~\ref{thm:abs-word}}

    Let $\theta_1, \ldots, \theta_m$
    be all the subformulas of the type 
    $p(t_1,\ldots,t_k)$ occurring in $\phi$,
    $n_i$ the maximum lookahead of $\Next$ in $\theta_i$,
    and $\prop_i$ the corresponding atomic proposition, as described in Section~\ref{sec:encoding}.
    Moreover, let $w' = \toltl(\phi,w)$.
    It is sufficient to show that, for all $a \in [ |w| ]$,
    and all $i \in [m]$,
    $w_{\geq a} \models \theta_i$ iff $w'_{\geq a} \models \Tom^{n_i} \prop_i$.
    
    First, assume that $w_{\geq a} \models \theta_i$.
    Since the maximum lookahead of $\Next$ in $\theta_i$ is $n_i$,
    the length of $w$ and $w'$ is at least $a + n_i$.
    So, the suffix $w'_{\geq a+n_i}$
    is well defined and contains at least one letter.
    Then, by definition of the abstract word $w'$, it holds $\prop_i \in w'_{a+n_i}$, as desired.
    
    Conversely, assume that $w'_{\geq a} \models \Tom^{n_i} \prop_i$.
    As before, this implies that the length of $w$ and $w'$
    is at least $a + n_i$.
    By definition, we have that $\prop_i \in w'_{\geq a + n_i}$
    and consequently that $w_{\geq a} \models \theta_i$,
    as desired.

\section{Proof of Theorem~\ref{thm:sdta2}} 
Let $\aut_\varphi$ be the \SDWA described in Section~\ref{sec:sdwa},
with intermediate steps $\varphi'$ and $\aut_{\varphi'}$.
First, we show that $\langW(\varphi) = \langW(\aut_\varphi)$.

Let $w$ be a data word in $\langW(\phi)$.
By Theorem~\ref{thm:abs-word},
the abstract word $w' = \toltl(\phi,w)$ on the alphabet 
$\{0,1\}^m$ satisfies $\phi'$ and therefore is accepted
by $\aut_{\phi'}$.
Let $n= |w| = |w'|$ and let $q_0,q_1,\ldots,q_n$ be the
(unique) accepting run of $\aut_{\phi'}$ on $w'$.
We build an accepting run $\bar{s} = s_0,s_1,\ldots,s_n$
of $\aut_\phi$ on $w$.
The $\mathit{state}$ component of $\bar{s}$ follows $q_0,q_1,\ldots,q_n$, and the $\mathit{data}$ components
start with default values and then progressively
store the last symbols read from $w$, as prescribed by the transition
predicate $\psi^\Delta$.
To show that $\bar{s}$ is a valid run of $\aut_{\phi}$,
it remains to prove that
condition \eqref{symbolic-delta} holds at every step of $\bar{s}$.
Let $j \in [n]$ be a step and $i \in [m]$ identify
the subformula $\theta_i$ of $\phi$.
Consider the data word 
$$
u \defeq s_{j-1}.\mathit{data}^{n_i} \cdot 
s_{j-1}.\mathit{data}^{n_i-1} \cdots 
s_{j-1}.\mathit{data}^{1} \cdot a \,.
$$
By construction, $\delta(q_{j-1}, w'_j) = q_j$.
Assume first that $j \leq n_i$.
Then, $\prop_i \in w'_j$ iff $\mathit{def}_{n_i - j+1} \cdot w \models \theta_i$.
Now, notice that $u$ coincides with the first $n_i+1$
symbols in $\mathit{def}_{n_i - j+1} \cdot w$.
Since $\theta_i$ does not depend on symbols beyond the first
$n_i +1$, we obtain that $u \models \theta_i$ iff
$\prop_i \in w'_j$, as required.

Next, assume that $j > n_i$.
Then, $\prop_i \in w'_j$ iff $w_{\geq j - n_i} \models \theta_i$.
As before, $u$ coincides with the first $n_i+1$
symbols in $w_{\geq j - n_i}$, and the thesis follows
from the same argument.

Conversely, let $w \in \langW(\aut_\varphi)$
and let $\bar{s} = s_0,s_1,\ldots,s_n$ be an accepting run
of $\aut_\varphi$ on $w$.
Let $q_j = s_j.\mathit{state}$, we prove that the sequence $\bar{q} = q_0,q_1,\ldots,q_n$ forms an accepting run of 
$\aut_{\varphi'}$ on $w' \defeq \toltl(\phi, w)$.
By construction, $q_0$ is the initial state of 
$\aut_{\varphi'}$ and $q_n$ is an accepting state.
By definition of run, for any step $j \in [n]$, it holds 
$\psi^\Delta(s_{j-1}, w_j, s_j)$.
By definition of $\psi^\Delta$,
$\delta(s_{j-1}.\mathit{state}, \sigma_j) = s_j.\mathit{state}$, for some $\sigma_j \in 2^\AP$.
This proves that $\bar{q}$ is an accepting run of
$\aut_{\phi'}$ on $\sigma_1,\ldots,\sigma_n$.
Next, we show that $\sigma_1,\ldots,\sigma_n$ is in fact $w'$.
Condition \eqref{symbolic-delta} connects the content
of each $\sigma_j$ with the truth of each subformula $\theta_i$ 
on the data word obtained by concatenating the symbols
stored in the data buffer of the symbolic automaton.
Since that buffer contains the last symbols read from
the data word $w$, we obtain a correspondence
with the definition of $\toltl(\phi, w)$, which proves
the claim that $\sigma_1,\ldots,\sigma_n = w'$.
So, $w' = \toltl(\phi, w)$ is accepted by $\aut_{\phi'}$.
In turn, this implies that $w'$ satisfies $\phi'$,
and, by Theorem~\ref{thm:abs-word}, $w$ satisfies $\phi$,
as required.

Next, we show that the \SDWA $\aut_\varphi$ so defined is deterministic.
By construction, the initial state is plainly unique.
Assume by contradiction that there exist a symbolic state $s$, a symbol $a$,
and two distinct symbolic states $s',s''$ such that 
both $\psi^\Delta(s, a, s')$ and $\psi^\Delta(s,a,s'')$ hold true.
By definition, the $\mathit{data}^i$ components of $s'$ and $s''$ are uniquely
determined by $s$ and $a$, for all $i \in [N]$.
So, $s.\mathit{data}^i = s''.\mathit{data}^i$.
Since we are assuming that $s'$ and $s''$ are distinct, we conclude that
$s'.\mathit{state} \neq s''.\mathit{state}$.
By construction, there exist $\sigma', \sigma'' \in 2^\AP$ such that
$\delta(s.\mathit{state}, \sigma') = s'.\mathit{state}$ and
$\delta(s.\mathit{state}, \sigma'') = s''.\mathit{state}$.
It follows that $\sigma' \neq \sigma''$.
Let $\prop_i \in \AP$ be an atomic proposition on which $\sigma'$ and $\sigma''$ differ.
By definition of $\psi^\Delta$, we obtain that the data constraint $\theta_i$ is both 
true and false when evaluated over the same tuple of data, which is a contradiction.
We conclude that $s' = s''$ and $\aut_\varphi$ is deterministic.

\section{Proof of Theorem~\ref{thm: monitoring precise}}

First, assume that $w$ is a data word whose monitoring state is \CS.
By definition, $w \models \phi$.
Then, the monitoring procedure simulates $\aut_\phi$ on $w$, leading to a state $q$ that satisfies 
$\psi^F(q)$, because $\langW(\aut_\phi) = \langW(\phi)$
(see Theorem~\ref{thm:sdta2}).

Next, consider the modified automaton $\aut_\phi(q,\overline{F})$.
Since $\aut_\phi$ is deterministic, we have that
$\langW(\aut_\phi(q,\overline{F}))$ is the complement of $\langW(\aut_\phi(q,F))$.
Since $w w' \not\models \phi$,
$\aut_\phi$ on $w w'$ reaches a state $q'$
that violates $\psi^F$, i.e., $\psi^F(q') = \mathit{false}$.
The unique run of $\aut_\phi$ on $w w'$ can then be split
into a prefix corresponding to $w$ that ends in state $q$ and a suffix that starts from $q$ and ends in $q'$. The second part of the run is also a run of
$\langW(\aut_\phi(q,\overline{F}))$ on $w'$.
This proves that $w' \in \langW(\aut_\phi(q,\overline{F})) \neq \emptyset$.

This argument holds also in reverse, showing that if the monitoring procedure returns \CS, then the monitoring state of $w$ is indeed \CS.

Next, assume that the monitoring state of $w$ is \PS.
We immediately obtain that $\psi^F(q) = \mathit{true}$,
where again $q$ is the state reached by $\aut_\phi$
after reading $w$.
Suppose by contradiction that $\langW(\aut_\phi(q,\overline{F}))$ is not empty,
and let $w'$ be a data word in that language.
Then, the automaton $\aut_\phi$ upon reading $w w'$
will reach $q$ and then it will end up in a state $q'$
such that $\psi^F(q') = \mathit{false}$.
By Theorem~\ref{thm:sdta2}, this implies that
$w w' \not\models \phi$, which is a contradiction.

The reverse implication follows the same argument,
and the remaining cases for monitoring states \CV and
\PV can be proved analogously.
\end{document}